\documentclass[12pt]{article}
\usepackage{booktabs}  
\usepackage{amsmath}
\usepackage{amsthm}
\usepackage{amssymb}
\usepackage{mathrsfs}
\usepackage{gensymb}
\usepackage{textcomp}  
\usepackage{xcolor}
\usepackage{lipsum}    
\usepackage{graphicx}
\usepackage{caption}
\usepackage{subcaption}
\usepackage{bm, bbm}
\RequirePackage{diagbox, adjustbox}
\usepackage{natbib}
\usepackage[colorlinks=true,linkcolor=black, citecolor=black, urlcolor=black]{hyperref}
\usepackage{multirow}
\usepackage{multicol}
\usepackage{pdflscape}
\usepackage{afterpage}
\usepackage{everypage}
\usepackage{natbib}
\usepackage{float}
\usepackage{url} 

\begin{document}

\def\spacingset#1{\renewcommand{\baselinestretch}%
{#1}\small\normalsize} \spacingset{1}


\begin{center}
{\Large {\bf Computationally Scalable Bayesian SPDE Modeling for Censored Spatial Responses}} 
\end{center}

\vspace{.1in} 

\begin{center}
{\large { Indranil Sahoo$^{1}$, Suman Majumder$^{2}$, Arnab Hazra$^{3}$, Ana G. Rappold$^{4}$, Dipankar Bandyopadhyay$^{5}$}} \\

\bigskip

{\normalsize {\it $^{1}$Department of Statistical Sciences and Operations Research, Virginia Commonwealth University \\
$^{2}$Department of Statistics, University of Missouri \\ 
$^{3}$Department of Mathematics and Statistics, Indian Institute of Technology Kanpur \\ 
$^{4}$United States Environmental Protection Agency \\ 
$^{5}$Department of Biostatistics, Virginia Commonwealth University}} \\
\end{center}

\vspace{.1in}
\baselineskip 18truept

\begin{abstract}

Observations of groundwater pollutants, such as arsenic or Perfluorooctane sulfonate (PFOS), are riddled with left censoring. These measurements have impact on the health and lifestyle of the populace. Left censoring of these spatially correlated observations are usually addressed by applying Gaussian processes (GPs), which have theoretical advantages. However, this comes with a challenging computational complexity of $\mathcal{O}(n^3)$, which is impractical for large datasets. Additionally, a sizable proportion of the data being left-censored creates further bottlenecks, since the likelihood computation now involves an intractable high-dimensional integral of the multivariate Gaussian density. In this article, we tackle these two problems simultaneously by approximating the GP with a Gaussian Markov random field (GMRF) approach that exploits an explicit link between a GP with Mat\'ern correlation function and a GMRF using stochastic partial differential equations (SPDEs). We introduce a GMRF-based measurement error into the model, which alleviates the likelihood computation for the censored data, drastically improving the speed of the model while maintaining admirable accuracy. Our approach demonstrates robustness and substantial computational scalability, compared to state-of-the-art methods for censored spatial responses across various simulation settings. Finally, the fit of this fully Bayesian model to the concentration of PFOS in groundwater available at 24,959 sites across California, where 46.62\% responses are censored, produces prediction surface and uncertainty quantification in real time, thereby substantiating the applicability and scalability of the proposed method. Code for implementation is made available via \texttt{GitHub}. 

\end{abstract}

\noindent%
\vfill

\newpage
\spacingset{1.45} 

\section{Introduction}

Analysis of censored data has been around in the statistical literature since the late 1900s regularly, with the earliest instance of a statistical analysis of censored data being in 1766 \citep{rosen1955problems}.  Measurements are often censored due to limitations of measuring instruments, physical inability to acquire data, human error, or similar.  

While most analyses on censored data focus on right-censoring \citep{hosmer2008applied}, some applications, such as analyzing the concentration of contaminants such as arsenic or per-and polyfluoroalkyl substances (PFAS) in groundwater call for utilizing methods involving left-censored data. This kind of censoring is prevalent in environmental monitoring and has applications in environmental and public health, epidemiology, hydrology, agriculture, and more. Typically these applications involve geostatistical data, where the measurements are censored because they fall below the minimum detection limit (MDL) of the measuring instrument. Early applications would remove censored observations or replace them with makeshift values such as MDL or MDL/2, or impute them with the mean or median of observed responses. Such ad-hoc imputations can result in biased estimates of the overall spatial variability, as demonstrated by \cite{fridley2007data}. 

Recent approaches for statistical inference of spatially distributed censored data overwhelmingly considered the Expectation-Maximization (EM) algorithm \citep{militino1999analyzing}. \cite{ordonez2018geostatistical} proposed an exact maximum likelihood (ML) estimation of model parameters under censoring, called `CensSpatial', using the Stochastic Approximation of the Expectation Maximization \citep[SAEM;][]{delyon1999convergence} algorithm. To tackle the computational complexities arising from censored likelihoods in correlated data, Monte Carlo approximations have been employed, both within the classical framework \citep{stein1992prediction, rathbun2006spatial}, and the Bayesian paradigm \citep{de2005bayesian, tadayon2017bayesian, sahoo2021contamination}. For example, \cite{schelin2014spatial} introduced a semi-naive approach that utilizes an iterative algorithm and variogram estimation to determine imputed values at locations where data are censored. Finally, various data augmentation techniques have been proposed to facilitate analysis of spatially correlated censored data \citep{abrahamsen2001kriging, hopke2001multiple, fridley2007data, sedda2012imputing}. However, the scalability of the suggested approaches is restricted, rendering them unsuitable for the analysis of large spatial datasets featuring censoring, a common occurrence in contemporary scientific research.

Gaussian processes \citep[GPs;][]{schulz2018tutorial} are heavily used for modeling continuous spatial data due to their several theoretical and computational advantages: the likelihood involves only the first two moments, conditional independence and zeros in the underlying precision matrix are equivalent, and various linear algebraic results are well-known in the literature that are required for computing covariance matrices \citep{gelfand2016spatial}. However, once the number of spatial sites is large and data at a large proportion of sites are censored, likelihoods based on the underlying GPs involve an intractable high-dimensional integral of a multivariate Gaussian density. This paper aims to overcome the computational challenges inherent to censored likelihoods for high-dimensional spatial settings through a combined application of two key steps:

\begin{enumerate}

    \item We focus on a fully Bayesian method for censored point referenced data, where the underlying GP is approximated as a Mat\'ern-like Gaussian Markov random field \citep[GMRF,][]{rue2005gaussian}. The GMRF is obtained as the solution of a stochastic partial differential equation \citep[SPDE,][]{lindgren2011explicit} on a fine mesh, which yields a sparse precision matrix of the underlying basis function coefficients. This sparse spatial structure then allows for fast and scalable Bayesian computations.

    \item We consider a GMRF-based measurement error model that incorporates a nugget effect in the formulation of the underlying GMRF, which expedites the imputation process for the censored observations. This inclusion effectively reduces the computational burden associated with censored likelihoods \citep{hazra2018semiparametric, yadav2019spatial, zhang2021hierarchical}.

\end{enumerate}

We draw inferences regarding model parameters using an adaptive Markov Chain Monte Carlo (MCMC) sampling approach, where we use random walk Metropolis-Hastings (MH) steps within Gibbs sampling. Extensive simulations demonstrate the scalability and performance of the proposed methodology in comparison to the `CensSpatial' algorithm, and the traditional local likelihood method \citep{wiens2020modeling, sahoo2023estimating} applied using Vecchia's approximation \citep{vecchia1988estimation}, across varying degrees of censoring and varying grid sizes. While the idea of a GMRF-based measurement error model has been explored in the context of spatial extremes \citep{hazra2021realistic, cisneros2023combined}, where replications of the underlying spatial processes are available and censoring a portion of the data is artificial, per our knowledge this modeling strategy has not been explored yet for high-dimensional censored spatial data without replications. Although, the lack of temporal replications typically leads to unstable computations, the proposed stable and scalable computational framework is specifically tailored for handling censored spatial data without requiring temporal replications. Furthermore, unlike previous studies involving GMRF-based measurement error model, a novel feature of the proposed approach is the inclusion of spatial predictions. 

PFAS constitute a substantial group of synthetic compounds absent in natural environments, notable for their resistance to heat, water, and oil. PFAS are persistent in the environment and can accumulate within the human body over time, and are toxic at relatively low concentrations \citep{wang2017never}. Exposure to elevated levels of PFAS can lead to various adverse health outcomes, including developmental issues during pregnancy, cancer, liver impairment, immune system dysfunction, thyroid disruption, and alterations in cholesterol levels. Due to their chemical robustness, PFAS endure in the environment, and are resistant to degradation. Contamination of drinking water with PFAS occurs through the use or accidental spillage of products containing these substances onto land or into waterbodies \citep{hepburn2019contamination}. PFAS also poses a significant threat to public health, with concentrations measured as high as 82 parts per trillion (ppt) in June 2023 in the United States, above the limit set by the United States Environmental Protection Agency (EPA) at 70 ppt \citep{cordner2019guideline}. A recent study \citep{andrews2020population} estimated that PFAS in publicly accessible drinking water could be affecting as many as 200 million people across the United States. Along similar lines, a robust Bayesian hierarchical approach was proposed \citep{smalling2023per} to accommodate left-censored PFAS responses; however, the model was implemented on a limited number of sample site locations. As such, the review of existing literature highlights the necessity for further investigation into PFAS occurrences in groundwater, alongside the development of fast and efficient approaches to handle large-scale left-censored spatial data in real time. Motivated by data on PFAS concentrations collected by the Groundwater Ambient Monitoring and Assessment (GAMA) program \citep{GAMA} across the state of California, we develop our Bayesian scalable model for spatially-referenced left-censored PFAS responses, in an attempt to provide a more accurate quantification of the groundwater contamination within the state. These data, collected by GAMA since 2019, allow thorough quality assessments of water sources and enable the establishment of safety thresholds for select PFAS constituents. Thus, our analysis is capable of identifying possible hotspots of higher PFAS concentration, thereby providing insights for further analysis of impacts on public health. 

The subsequent sections of the paper are organized as follows. In Section \ref{data}, we provide details regarding the dataset on the groundwater levels of PFAS within the state of California, along with some exploratory analyses. We outline our methodology and related computational details in Section \ref{methodology}, and test its scalability and predictive performance on simulated datasets in Section \ref{simulation}. In Section \ref{application}, we apply the proposed methodology to the PFAS dataset and report the findings. We conclude with a brief discussion in Section \ref{discussion}.

\section{Motivating PFAS data}
\label{data}

The groundwater PFAS data for the state of California are available online at the website \href{https://gamagroundwater.waterboards.ca.gov/gama/datadownload}{GAMA Groundwater Information Systems} under the label \texttt{Statewide PFOS Data}. In this paper, we focus specifically on the measurements of the chemical substance, known as Perfluorooctane sulfonate (PFOS). The dataset contains 24,959 measurements (in ng/L) of PFOS concentration and their locations (in longitudes and latitudes) as well as indicators of whether the observations are censored or not and the corresponding censoring limits within the state of California. Almost half of the measurements (46.62\%) are censored observations, with varying degrees of censoring limits. 

\begin{figure}
\centering
    \includegraphics[width=0.8\linewidth]{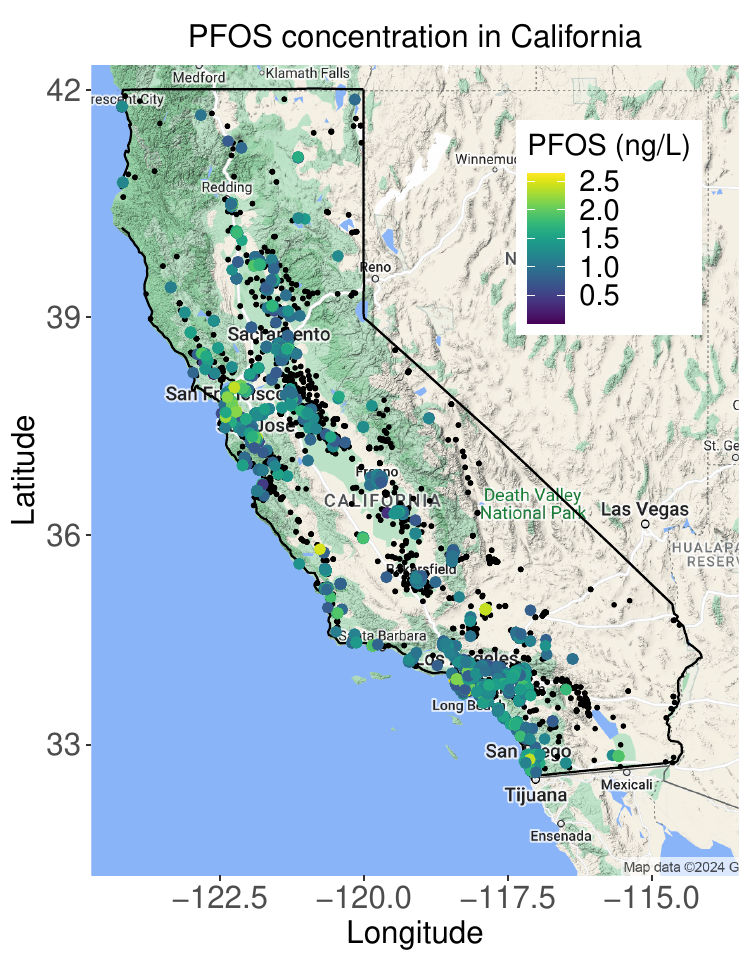}
    \caption{Concentrations of (transformed) PFOS, using the transformation $g(\textrm{PFOS}) = \log(1 + \log(1 + \textrm{PFOS}))$, measured at 24,959 irregularly-sampled spatial locations across the state of California (in ng/L). The tiny black dots indicate the sites with censored data.} 
    \label{all_locs}
\end{figure}

Figure \ref{all_locs} shows transformed PFOS concentration measurements, after transforming the raw PFOS by $g(\textrm{PFOS}) = \log(1 + \log(1 + \textrm{PFOS}))$ at the 24,959 irregularly sampled spatial locations across the state of California, prompting an approximate spatial inference model to be employed, which also accounts for the huge proportion of censored observations. Most observation sites are towards the densely populated areas on the coast. The censored observations are presented as tiny black dots in Fig \ref{all_locs}. The censored observations are all over the spatial domain, and are not limited  to one single area. Most of the measurements are below 150 ng/L, but there are observations as high as 1,330,000 ng/L, which is well-beyond the safety limit prescribed by the EPA.

\begin{figure*}
\centering
    \includegraphics[width=0.8\textwidth]{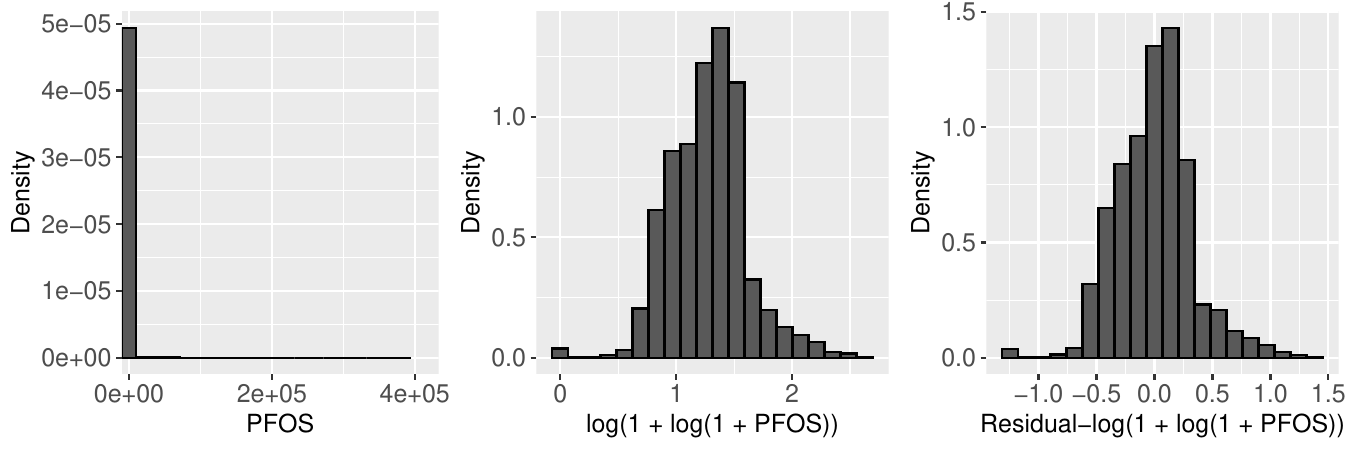}
    \caption{Pictoral representation of the raw PFOS responses. \emph{Left panel:} Histogram of raw concentrations of PFOS across sites where the data are not censored. \emph{Middle:} Histogram of non-censored PFOS concentrations after the transformation $g(\textrm{PFOS}) = \log(1 + \log(1 + \textrm{PFOS}))$. \emph{Right panel:} Histogram of the residuals obtained after regressing non-censored transformed PFOS observations to longitude and latitude, via a linear model.} 
    \label{histograms}
\end{figure*}

The histogram of the raw non-censored PFOS observations is presented in the left panel of Figure \ref{histograms}. The raw data exhibit a highly positively skewed nature, and thus a stationary Gaussian process assumption naturally becomes questionable, even after considering a spatially-smooth mean surface with covariates, like longitude and latitude. Following an exploration of different transformations of the raw data such that the histogram behaves in an approximately bell-shaped fashion, we identify that the iterated log-transformation $g(\textrm{PFOS}) = \log(1 + \log(1 + \textrm{PFOS}))$ performs reasonably well; the histogram of the transformed PFOS data is presented in the middle panel of Figure \ref{histograms}. We further explore the effects of the natural covariates longitude and latitude on the transformed data; following a simple linear regression, we obtain the residuals, and their histogram is presented in the right panel of Figure \ref{histograms}. This histogram is reasonably bell-shaped and thus we model this transformed PFOS data using a Gaussian process framework with a regression structure for the mean process where we allow longitude and latitude as covariates.

\begin{figure}
\centering
    \includegraphics[width=0.8\linewidth]{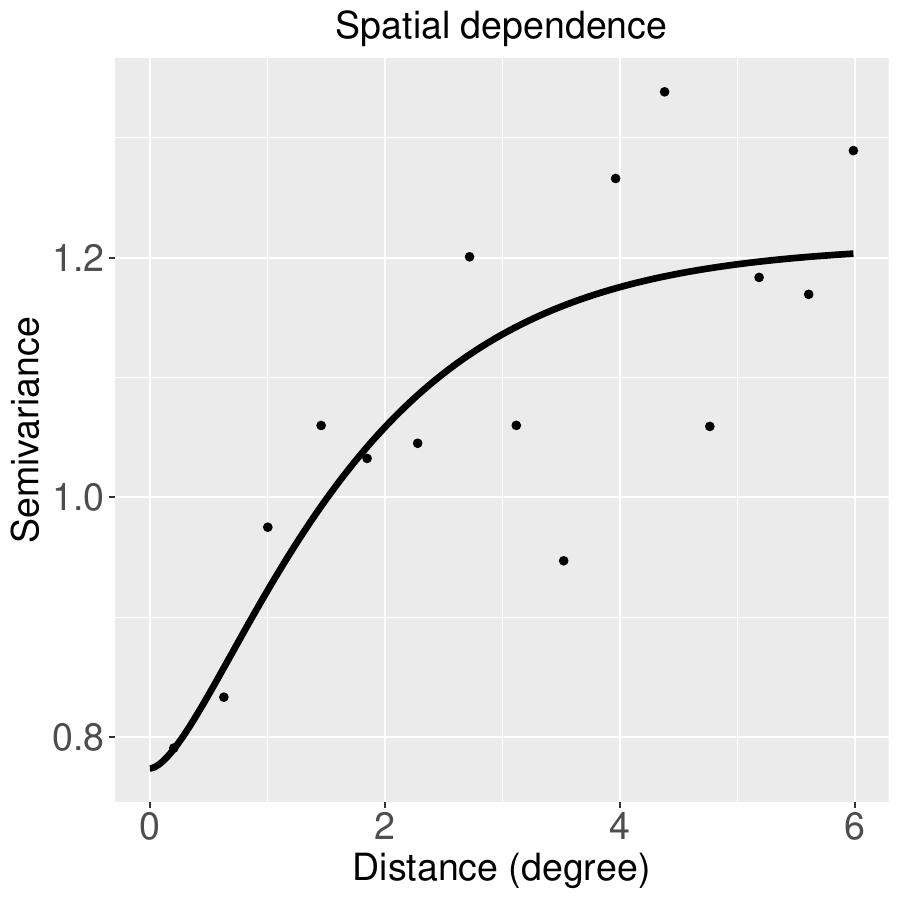}
    \caption{Sample semivariogram of residuals obtained after regressing non-censored transformed PFOS observations to longitude and latitude, as a function of distance (dots). The overlapped solid line represents the fitted population semivariance obtained from \eqref{matcor}.}
    \label{variogram}
\end{figure}

We further explore the presence of spatial correlation following a variogram analysis of the residuals (scaled by their sample standard deviation) discussed above. The sample semivariogram at distance $d$ is defined as 
$$
\widehat{\gamma}(d) = \frac{1}{2N(d)}\sum_{i = 1}^n \sum_{j = 1}^{i} w_{ij}(d)(R(\bm{s}_i) - R(\bm{s}_j))^2,
$$
where, $R(\bm{s}_i)$ and $R(\bm{s}_j)$ are the residuals at spatial sites $\bm{s}_i$ and $\bm{s}_j$, $w_{ij}(d) = 1$ if $d_{ij} \in (d - h, d + h)$ and $w_{ij} = 0$ otherwise, $d_{ij}$ being the distance between $\bm{s}_i$ and $\bm{s}_j$. Also, $N(d)$ is the number of pairs with $w_{ij}(d) = 1$. The sample semivariogram, presented in Figure \ref{variogram}, indicates the presence of spatial correlation, along with the presence of possible nugget effects \citep{bivand2008applied}. We fit an isotropic Mat\'ern spatial correlation function, with its smoothness parameter set to one, plus a nugget effect, given by
\begin{equation}
\label{matcor}
    \rho(\bm{s}_i, \bm{s}_j) \equiv \rho(d) = \gamma\frac{d}{\phi} \kappa_{1}\left( \frac{d}{\phi} \right) + (1 - \gamma) \mathbbm{1}(\bm{s}_i = \bm{s}_j),
\end{equation}
where, $d$ is the Euclidean distance between locations $\bm{s}_i$ and $\bm{s}_j, \phi > 0$ is the range parameter, $\gamma \in [0, 1]$ is the ratio of spatial to total variation, $\kappa_1(\cdot)$ is the modified Bessel function of second kind with degree 1, and $\mathbbm{1}(\cdot)$ is the indicator function. The fitted population semivariance indicates a reasonable fit to the sample semivariogram. While these exploratory analyses are based on non-censored observations only, they indicate a need for proper spatial modeling after considering the censored nature of a large proportion of the data. Specifically, most of the observations near the eastern regions of the study domain are censored, and ignoring them in the spatial prediction would lead to poor spatial prediction for the nearby regions.

\section{Methodology}
\label{methodology}

Let $Y(\bm{s})$ represent transformed PFOS concentration at a spatial location $\bm{s} \in \mathcal{D} \subset \mathbb{R}^2$, where $\mathcal{D}$ represents the spatial domain of interest, i.e., the entire state of California in our case. We model $Y(\bm{s})$ as 
$$
Y(\bm{s}) = \bm{X}(\bm{s})^T\bm{\beta} + \tau^{-1/2} Z(\bm{s}),
$$
where, $X(\bm{s}) = [X_1(\bm{s}), \ldots, X_p(\bm{s})]^T$ denotes the vector of $p$ covariates at location $\bm{s}$, $\bm{\beta} = [\beta_1, \ldots, \beta_p]^T$ is a vector of unknown regression parameters and $\tau > 0$ is a spatially-constant precision parameter. Given the absence of meaningful covariates in our data, we choose $\bm{X}(\bm{s}) = [1, \textrm{longitude}(\bm{s}), \textrm{latitude}(\bm{s})]^T$ for our analysis. We assume that $Z(\cdot)$ is a standard (mean zero and variance one at each site) GP with an isotropic Mat\'ern spatial correlation plus a nugget effect, given by \eqref{matcor}. While the incorporation of the nugget effect is justified by our exploratory analysis (nonzero semivariance at origin), it effectively addresses the issue of censoring in the response, thereby circumnavigating the computational burden occurring due to censored likelihoods \citep{hazra2018semiparametric, yadav2019spatial, zhang2021hierarchical}. Here, we fix the smoothness parameter of the Mat\'ern correlation of the purely spatial component of \eqref{matcor} to one. In practice, it is difficult to estimate the smoothness parameter from the data, and hence it is generally fixed. Besides, we later build a stochastic partial differential equation-based approximation of the Mat\'ern correlation structure, where fixing the smoothness parameter to one is a standard choice \citep{hazra2021realistic, cisneros2023combined}.

Suppose the data are observed (either censored or non-censored) at the set of sites $\mathcal{S} = \{ \bm{s}_1, \ldots, \bm{s}_n \}$. In matrix notations, the spatial linear model can be written as 
\begin{equation}
\label{model0}
    \bm{Y} = \bm{X}\bm{\beta} + \tau^{-1/2} \bm{Z},
\end{equation}
where, $\bm{Y}_{(n \times 1)}$ is the response vector, $X_{(n \times p)}$ is the matrix of covariates, $\bm{\beta}_{(p \times 1)}$ is the vector of regression coefficients, and $\bm{Z}_{(n \times 1)} \sim \textrm{MVN}(\bm{0}, \gamma \bm{\Sigma} + (1 - \gamma) \bm{I}_n)$, where $\bm{\Sigma}$ is the Mat\'ern correlation matrix, and $\bm{I}_n$ denotes the identity matrix of order $n$. Here, by construction and the PFAS dataset, $\bm{\Sigma}$ is non-singular and $\bm{X}$ has full rank.

In a spatial censored linear (SCL) model, it is further assumed that $Y(\bm{s})$ is not fully observed at all spatial locations. Motivated by the dataset considered, we assume $Y(\cdot)$ to be left-censored at sites $\mathcal{S}^{(c)} = \{ \bm{s}^{(c)}_1, \ldots, \bm{s}^{(c)}_{n_c} \} \subset \mathcal{S}$ and the corresponding censoring levels be $\mathcal{U} = \{ u_1, \ldots, u_{n_c} \}$. However, a similar approach can be applied if the response is right or interval-censored. We define the censoring indicator $\delta(\bm{s})$ as 
$$
\delta(\bm{s}) = \begin{cases}
    1, & \text{if $Y(\bm{s})$ is censored at site $\bm{s}$}, \\
    0, & \text{otherwise},
\end{cases} 
$$
and the vector of censored observations as $\bm{v} = [ Y(\bm{s}_i): \delta(\bm{s}_i) = 1]^T \equiv [Y(\bm{s}^{(c)}_1), \ldots, Y(\bm{s}^{(c)}_{n_c})]^T.$ Then, for censored spatial data, the likelihood is given by
\begin{equation} \label{likelihood}
    \mathcal{L}(\bm{\theta}) = \int_{\bm{v} \leq \bm{u}} f_{\textrm{MVN}}(\bm{y}; \bm{X} \bm{\beta}, \tau^{-1} [\gamma \bm{\Sigma} + (1 - \gamma) \bm{I}_{n}]) \hspace{0.05cm} d\bm{v},
\end{equation}
where the integral is over the censored responses $\lbrace \bm{y}: y(\bm{s}_i) \leq u_i ~\textrm{if}~ \bm{s}_i \in  \mathcal{S}^{(c)} \rbrace$ and $f_{\textrm{MVN}}(\cdot; \bm{\mu}, \bm{\Sigma})$ denotes a multivariate normal density with mean $\bm{\mu}$ and covariance matrix $\bm{\Sigma}$. A version of this likelihood has been studied in \cite{sahoo2021contamination}.

\subsection{Approximation of the Mat\'ern Gaussian Process}
\label{appGP}

Here, we follow the approximation strategy of the Mat\'ern GP with nugget \citep{hazra2021realistic}. To ensure computational efficiency, we choose to approximate the Gaussian process $Z(\cdot)$ with a GP $\tilde{Z}(\cdot)$, constructed from a Gaussian Markov random field (GMRF) defined on a finite mesh, thereby circumventing the computational overhead associated with the dense correlation matrix inherent in the exact Mat\'ern GP defined by \eqref{matcor}. This strategy capitalizes on the direct correspondence between continuous-space Mat\'ern GP with dense covariance matrices and GMRFs with sparse precision matrices \citep{lindgren2011explicit}, which yields an approximate data process
$$
\tilde{Y}(\bm{s}) = \bm{X}(\bm{s})^T\bm{\beta} + \tau^{-1/2} \tilde{Z}(\bm{s}), ~~ \bm{s} \in \mathcal{D}. 
$$

\begin{figure*}
\centering
\includegraphics[height = 0.35\linewidth]{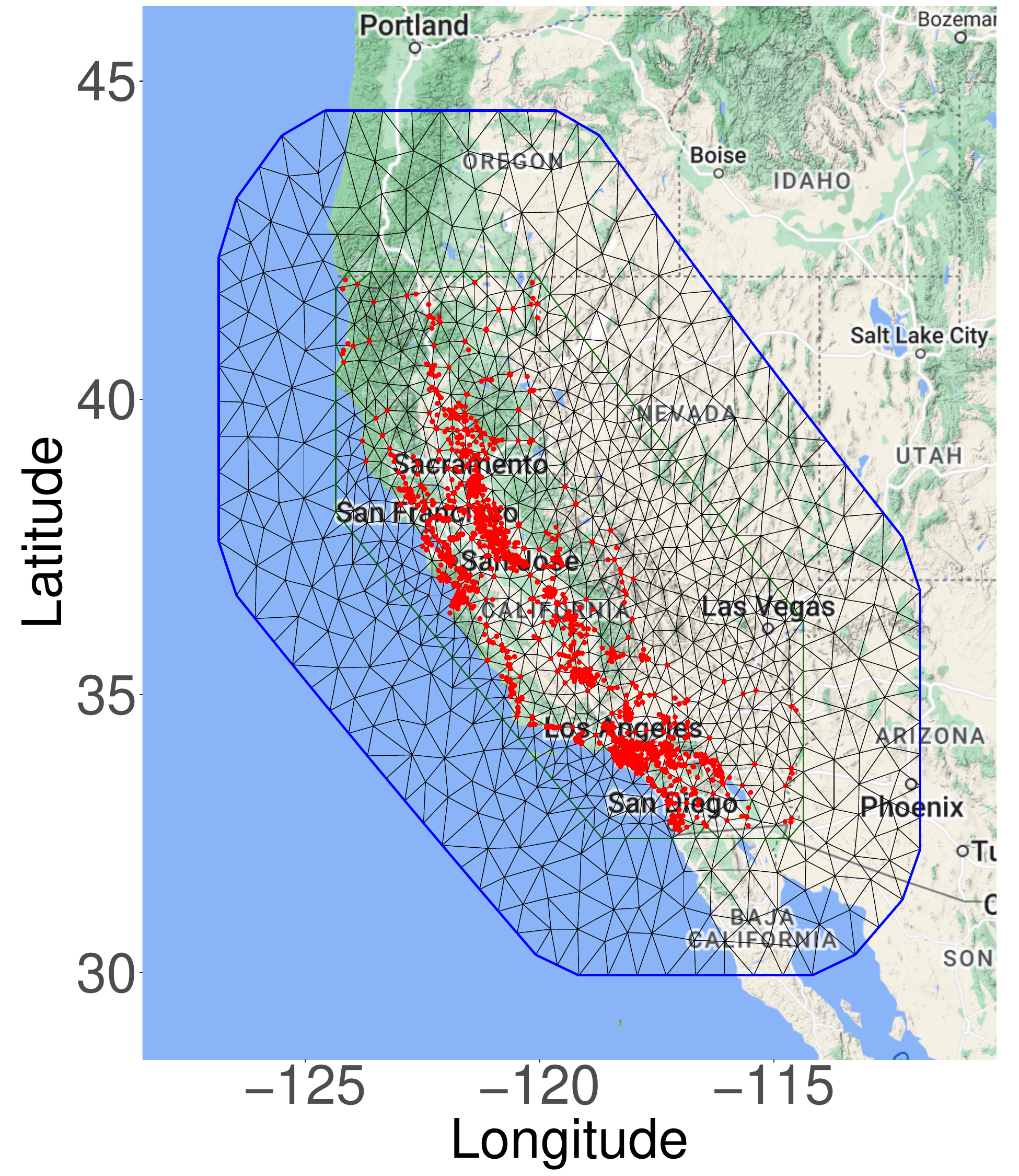} 
\includegraphics[height = 0.35\linewidth]{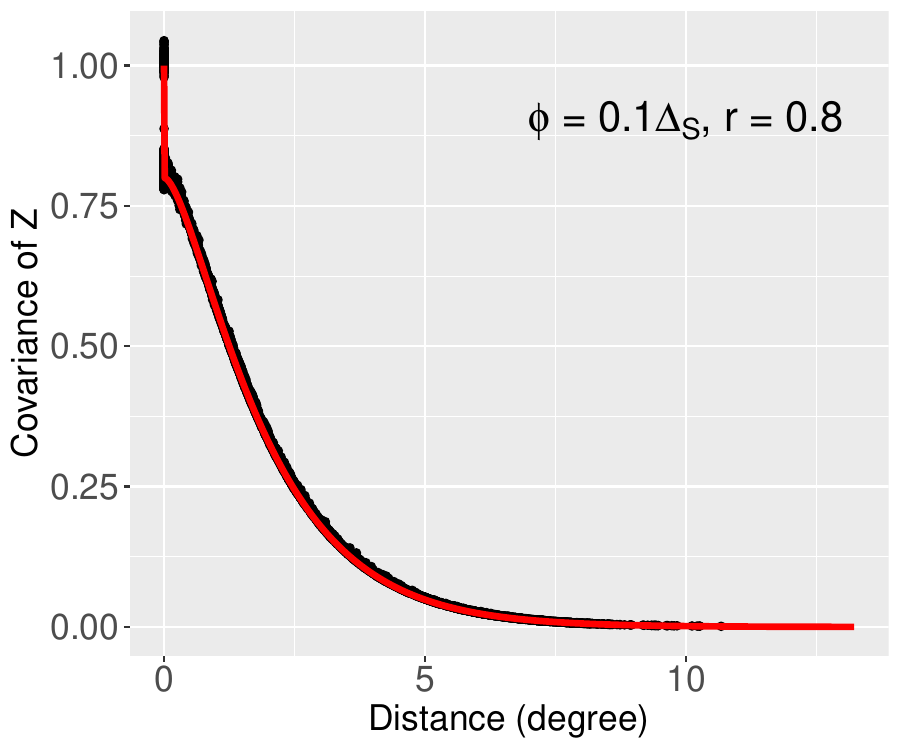}
\vspace{-2mm}
\caption{\emph{Left:} Triangulated mesh over California, that is used to approximate the spatial SPDE process $Z(\cdot)$. \emph{Right:} Comparison of the true Mat\'ern correlation (solid line) and the pairwise covariances between two spatial locations obtained by the SPDE approximation (points), as a function of distance. The parameters are set to $\phi = 0.1\Delta_{\mathcal{S}}$, with $\Delta_{\mathcal{S}}$ being the maximum spatial distance between two locations in the domain, and $r = 0.8$.}
\label{spde_approximation}
\end{figure*}

For $\gamma = 1$, the Gaussian Mat\'ern process $Z(\cdot)$ can be obtained as a solution to the linear Stochastic Partial Differential Equation (SPDE)
\begin{equation}
\label{SPDE}
(\phi^{-2} - \Delta)Z(\bm{s}) = 4\pi \phi^{-2}\mathcal{W}(\bm{s}),  \hspace{0.5cm} \bm{s} \in \mathbb{R}^2,
\end{equation}
where $\mathcal{W}(\cdot)$ is a Gaussian white noise process, and $\Delta$ is the Laplacian. The solution $Z(\bm{s})$ to the SPDE can be effectively approximated through finite-element methods \citep{ciarlet2002finite} applied to a triangulated mesh defined within a bounded region of $\mathbb{R}^2$, where the triangulation is formed through a refined Delaunay triangulation process \citep{borouchaki1995fast}. In practical applications, the mesh can be easily constructed using the (currently depreciated) \texttt{inla.mesh.2d} function, implemented in the \texttt{R} package \texttt{INLA} (\url{www.r-inla.org}) or the \texttt{fm\_mesh\_2d\_inla} function in the \texttt{R} package \texttt{fmesher} (\url{https://cran.r-project.org/package=fmesher}); see \cite{lindgren2015bayesian} for more details. The left panel of Figure \ref{spde_approximation} depicts the mesh utilized in the data application discussed in Section \ref{application}.

Let $\mathcal{S}^* = \lbrace \bm{s}_1^*, \ldots, \bm{s}_N^*\rbrace$ denote the set of mesh nodes. We construct a finite-element solution by writing $\Tilde{Z}(\bm{s}) = \sum_{j = 1}^N \zeta_j(\bm{s}) Z_j^*$, and plugging it in \eqref{SPDE} in place of $Z(\cdot)$. Here, $\lbrace \zeta_j(\cdot)\rbrace$ are piecewise linear and compactly-supported ``hat'' basis functions defined over the mesh, and $\lbrace Z_j^* \rbrace$ are normally distributed weights defined for each basis function (that is, one for each mesh node in $\mathcal{S}^*$). Then, $\bm{Z}^* = [Z_1^*, \ldots, Z_N^*]^T \sim \textrm{MVN}(\bm{0}, \bm{Q}_{\phi}^{-1})$, where the $(N \times N)$-dimensional precision matrix $\bm{Q}_{\phi}$ can be written as
\begin{equation}
    \bm{Q}_{\phi} = \frac{\phi^2}{4\pi} \left[ \frac{1}{\phi^4}\bm{D} + \frac{2}{\phi^2} \bm{G}_1+ \bm{G}_2 \right],
\end{equation}
where $\bm{D}, \bm{G}_1$, and $\bm{G}_2$ are sparse ($N \times N$)-dimensional finite-element matrices that can be obtained as follows. The matrix $\bm{D}$ is diagonal with its $j^{th}$ diagonal entry $D_{j, j} = \langle \zeta_j(\cdot), 1 \rangle$, where $\langle f, g \rangle = \int f(\bm{s})g(\bm{s})d\bm{s}$ denotes an inner product. Similarly, $\bm{G}_1$ has the elements $G_{1; j, k} = \langle \nabla \zeta_j(\cdot), \nabla \zeta_k(\cdot)\rangle$ and $\bm{G}_2 = \bm{G}_1\bm{D}^{-1}\bm{G}_1$.  Efficient computation of these sparse matrices is implemented using the function \texttt{inla.mesh.fem} from the \texttt{R} package \texttt{INLA}. For further theoretical details, see \cite{bakka2018spatial} and \cite{lindgren2022spde}.

In order to map the spatial random effects $\bm{Z}^*$ (defined across mesh nodes) back to the observation locations $\mathcal{S}$, we use an $(n \times N)$-dimensional projection matrix $\bm{A}$. The $(i,j)^{th}$ element of this matrix corresponds to $\zeta_j(\bm{s}_i)$ for every spatial location $\bm{s}_i \in \mathcal{S}$ and mesh node $\bm{s}_j^* \in \mathcal{S}^*$, allowing us to compute $\bm{A}\bm{Z}^*$, the projection of $\bm{Z}^*$ at the data locations. The generation of the matrix $\bm{A}$ is carried out through the function \texttt{inla.spde.make.A} within the \texttt{R} package \texttt{INLA}.
  
In presence of a nugget effect $\gamma \in [0, 1]$ in \eqref{matcor}, the model for $\Tilde{\bm{Z}} = [\Tilde{Z}(\bm{s}_1), \ldots, \Tilde{Z}(\bm{s}_n)]^T \equiv [\Tilde{Z}_1, \ldots, \Tilde{Z}_n]^T$ becomes
$$
\Tilde{\bm{Z}} = \sqrt{\gamma} \bm{A}\bm{Z}^* + \sqrt{1 - \gamma} \bm{\eta},
$$
where $\bm{\eta} = [\eta_1, \ldots, \eta_n]^T$ with $\eta_i \overset{IID}{\sim} \textrm{N}(0, 1), i = 1, \ldots, n$. Then, the covariance matrix of $\Tilde{\bm{Z}}$ can be written as 
\begin{equation}
\label{appmat}
\bm{\Sigma}_{\Tilde{\bm{Z}}} = \gamma \bm{A}\bm{Q}_{\phi}^{-1}\bm{A}^T + (1 - \gamma)\bm{I}_n.
\end{equation}

The SPDE approach yields a precise approximation of the true correlation structure; see the right panel of Figure \ref{spde_approximation}. Leveraging the sparsity of the matrix $\bm{Q}_{\phi}^{-1}$, we facilitate rapid Bayesian computations. Furthermore, we exploit the conditional independence structure of $\Tilde{\bm{Z}}| \bm{Z}^* \sim \textrm{MVN}(\sqrt{\gamma}\bm{A}\bm{Z}^*, (1 - \gamma)\bm{I}_n)$ for swift imputation of left-censored observations.

\subsection{Final hierarchical model}\label{final}
Suppose we write the vector of the final approximate data process, $\tilde{Y}(\cdot)$, evaluated at $\mathcal{S}$ by $\tilde{\bm{Y}}=[\tilde{Y}(\bm{s}_1), \ldots, \tilde{Y}(\bm{s}_n)]^T$. We define a rescaled random effects vector defined at mesh nodes by $\tilde{\bm{Z}}^* = \sqrt{\gamma / \tau} \bm{Z}^*$. Then, similar to \eqref{model0}, the model for $\tilde{\bm{Y}}$ can be written hierarchically as
\begin{eqnarray} \label{hierarchical}
\nonumber  \tilde{\bm{Y}} | \tilde{\bm{Z}}^* &\sim& \textrm{MVN}\left(\bm{X}\bm{\beta} + \bm{A}  \tilde{\bm{Z}}^*, \tau^{-1} (1 - \gamma) \bm{I}_n \right), \\
\nonumber  \tilde{\bm{Z}}^* &\sim& \textrm{MVN}(\bm{0}, \gamma \tau^{-1} \bm{Q}_{\phi}^{-1}), \\
    \{ \bm{\beta}, \tau, \phi, \gamma \} &\sim& \pi(\bm{\beta}) \times \pi(\tau) \times \pi(\phi) \times \pi(\gamma).
\end{eqnarray}
Here the last layer of the model indicates prior choices for the model parameters that we discuss in Section \ref{computation}. Instead of the likelihood based on \eqref{model0}, we fit the approximate data process \eqref{hierarchical} to the actual observation process.

\subsection{Prediction}\label{Prediction}

Let $\mathcal{S}^{(0)} = \lbrace \bm{s}^{(0)}_1, \ldots, \bm{s}^{(0)}_{n_0} \rbrace \subset \mathcal{D}$ denote a set of $n_0$ prediction sites, and define $\tilde{\bm{Y}}^{(0)} = [\tilde{Y}(\bm{s}^{(0)}_1), \ldots, \tilde{Y}(\bm{s}^{(0)}_{n_0})]^T$. Also, let $\bm{X}^{(0)}$ denote the $(n_0 \times p)$-dimensional design matrix, with its $i^{th}$ row $\bm{X}(\bm{s}^{(0)}_i), i = 1, \ldots, n_0$ denoting the vector of covariates at prediction location $\bm{s}^{(0)}_i$. For mapping the (scaled) spatial random effects $\tilde{\bm{Z}}^*$ (defined across mesh nodes) to the prediction locations $\mathcal{S}^{(0)}$, we use an $(n_0 \times N)$-dimensional projection matrix $\bm{A}^{(0)}$. The $(i,j)^{th}$ element of this matrix corresponds to $\zeta_j(\bm{s}^{(0)}_i)$ for every spatial location $\bm{s}^{(0)}_i \in \mathcal{S}^{(0)}$ and mesh node $\bm{s}_j^* \in \mathcal{S}^*$, allowing us to compute $\bm{A}^{(0)}\tilde{\bm{Z}}^*$, the projection of $\tilde{\bm{Z}}^*$ at $\mathcal{S}^{(0)}$. Then, given $\tilde{\bm{Z}}^*$, the conditional distribution of $\tilde{\bm{Y}}^{(0)}$ is
\begin{equation}
    \tilde{\bm{Y}}^{(0)} | \tilde{\bm{Z}}^* \sim \textrm{MVN}\left(\bm{X}^{(0)}\bm{\beta} + \bm{A}^{(0)}  \tilde{\bm{Z}}^*, \tau^{-1} (1 - \gamma) \bm{I}_n \right)
\end{equation}

\subsection{Computational details}
\label{computation}

Inference concerning the model parameters is conducted through Markov chain Monte Carlo (MCMC) sampling, implemented in \texttt{R}. Given the computational dependence on prior selections for the model parameters, we first specify these priors. Whenever feasible, we opt for conjugate priors and employ Gibbs sampling to update them iteratively. When prior conjugacy is unavailable, we resort to random walk Metropolis-Hastings (MH) steps for parameter updates. During the burn-in period, we adjust the candidate distributions within the MH steps to ensure that the acceptance rate throughout the post-burn-in period remains within the range of 0.3 to 0.5.

Here we draw samples from the full posterior $$\pi(\bm{\beta}, \tau, \phi, \gamma, \tilde{\bm{Z}}^*, \tilde{\bm{Y}}^{(c)} | \tilde{\bm{Y}}^{(nc)}),$$ 
where $\tilde{\bm{Y}}^{(c)}$ is the vector of censored data vector and $\tilde{\bm{Y}}^{(nc)}$ is the vector of non-censored data vector. For the vector of regression coefficients $\bm{\beta}$, we consider weakly-informative conjugate prior $\bm{\beta} \sim \textrm{MVN}(\bm{0}, 100^2 \bm{I}_p)$. The full conditional posterior of $\bm{\beta}$ is then multivariate normal and hence updated using direct sampling within Gibbs steps. Due to the strong posterior correlation between $\bm{\beta}$ and $\tilde{\bm{Z}}^*$, they are updated jointly within each Gibbs sampling step. For the hyperparameters involved in the correlation function in (\ref{matcor}), we consider the non-informative priors as well. Specifically, we choose $\phi \sim \textrm{Uniform}(0, 0.5\Delta_{\mathcal{S}})$ for the spatial range parameter, where $\Delta_{\mathcal{S}}$ is the largest Euclidean distance between two data locations, and $\gamma \sim \textrm{Uniform}(0, 1)$ for the nugget effect $\gamma$. We further designate a non-informative conjugate prior for the spatially-constant precision parameter $\tau$ in the process model, namely $\tau \sim \textrm{Gamma}(0.1, 0.1)$. The full conditional posterior distribution of $\tilde{\bm{Y}}^{(c)}$ is $\textrm{MVN}\left(\bm{X}^{(c)}\bm{\beta} + \bm{A}^{(c)}  \tilde{\bm{Z}}^*, \tau^{-1} (1 - \gamma) \bm{I}_n \right)$, where $\bm{X}^{(c)}$ and $\bm{A}^{(c)}$ are design matrix and SPDE projection matrix (from mesh nodes), respectively, corresponding to the locations with censored data, i.e., they comprise of the rows of $\bm{X}$ and $\bm{A}$ that correspond to the censored entries of $\bm{Y}$. 

\subsection{Software}

We have developed an open-source \texttt{R} package, called \texttt{CensSpBayes}, which implements the proposed approximate Mat\'ern GP model for large left-censored spatial data. Implementation code, along with details of execution using simulated data are made available at \url{https://github.com/SumanM47/CensSpBayes}.

\section{Simulation Study}
\label{simulation}

In this section, we conduct simulation studies using synthetic data to assess the efficacy of our proposed scalable modeling framework in terms of spatial prediction, while imputing censored values. We simulate 100 datasets over grids $\mathcal{D}^* = \{(i,j): i, j\in \{1/K, 2/K,\ldots, 1 \}\}$ of varying sizes within a spatial domain $[0, 1]^2$. We consider $K\times K$ grids with $K=20, 50, 100$, and 200, to demonstrate the computational power and scalability inherent to our proposed methodology.

For simulating the datasets, we consider an intercept term only and no other covariates, and we assume the true value of the regression coefficient to be $\beta^{\mbox{true}} = 5$. The true value of the range parameter of the spatial Mat\'ern correlation is chosen to be $\phi^{\mbox{true}} = 0.15 \times \Delta^*$, where $\Delta^* = \sqrt{2}$, the maximum spatial distance between two locations in $[0, 1]^2$. The smoothness parameter is set to one and not estimated while fitting our proposed model. The true ratio of partial sill to total variation is chosen to be $\gamma^{\mbox{true}} = 0.9$, and the true precision parameter is chosen to be $\tau^{\mbox{true}} = 1/5$. Exact simulations are conducted to generate datasets from a GP with Mat\'ern correlation, as given in \eqref{matcor}. 

Once the datasets are generated, we divide each dataset randomly into 80\% training and 20\% test datasets. Within each training set, we consider two different levels of censoring (denoted by L1 and L2) for the response by setting different values of the minimum detection limit (MDL):
\begin{itemize}
    \item[L1] Low censoring: The MDL is at the $15^{\mbox{th}}$ percentile point of observations and thus 15\% data are censored. 
    \item[L2] High censoring: The MDL is at the $45^{\mbox{th}}$ percentile point of observations and thus 45\% data are censored. 
\end{itemize}

For each of the two levels of censoring, we implement our proposed approximate Mat\'ern GP model under three different settings, denoted by S1, S2, and S3:

\begin{itemize}
    \item[S1] We selectively exclude spatial locations where observations are censored and apply the spatial model approximated via SPDE, as elaborated in Section \ref{appGP}, exclusively to the observed locations. This does not require any imputation of the censored observations. 
    \item[S2] We set the censored observations at the MDL and employ the SPDE-approximated Mat\'ern GP model, as detailed in Section \ref{appGP}. This once again circumvents the need for imputing the censored observations.
    \item[S3] We fit the full proposed model, treating the observations below MDL as censored observations, and implement the SPDE-approximated Mat\'ern GP model, as in Section \ref{appGP}, while simultaneously performing imputations for the censored observations.
\end{itemize}

In each of the three scenarios, the approximated spatial process using SPDE has been fitted to assess the effects of removing or considering ad hoc imputations of censored observations, in terms of mean squared prediction error (MSPE), in contrast to treating them as genuinely censored. The prior distributions for $\beta$ and $\gamma$ as described in Section \ref{computation} remain unchanged in the simulation study. However, for the range parameter, we assume $\phi \sim \textrm{Uniform}(0, 0.25\Delta^*)$.

For comparison, we consider two additional models, S4 and S5: 

\begin{itemize} 
    \item[S4] We use ML estimation to locally fit the full Matérn Gaussian censored likelihood, using Vecchia's approximation with $m = 30$ nearest neighbors, thereby estimating the model parameters under censoring. Here, we fix the Matérn smoothness parameter to 1 in keeping with the data generation mechanism.
    \item[S5] We use the `CensSpatial' algorithm to perform an exact ML estimation of model parameters, which implements the SAEM algorithm of \cite{ordonez2018geostatistical}. 
\end{itemize}

\begin{table*}[h]
    \caption{Average mean squared prediction error (MSPE) corresponding to model fitting under the five different settings (S1-S5) to simulated test data from a Mat\'ern GP that varies with censoring levels L1 (low-censoring; 15\%) and L2 (high-censoring; 45\%), and grid sizes. The values in parenthesis represent the corresponding average prediction standard errors. The lowest MSPE in each row is in bold. Since model S5 (`CensSpatial') was infeasible for larger grids, table entries (MSPE and standard errors) appear as `-'.}
    \label{table1}
        \centering
    \begin{tabular}{ccccccc}
    \hline
         Censoring& Grid & \multicolumn{4}{c}{Mean Squared Prediction Error}\\
         \cline{3-7}
         Level& size& S1 & S2 & S3 & S4 & S5\\ 
         \hline
         \multirow{4}{*}{L1}& $20 \times 20$ & 1.13(0.82)& 2.10(0.86)& \textbf{0.79}(0.85)& 99.15(6.0E18) & 0.89(0.97)\\
         & $50 \times 50$ & 0.87(0.74)& 2.03(0.80)& \textbf{0.61}(0.77)& 136.55(3.3E84) & -\\
         & $100 \times 100$ & 0.80(0.70)& 1.88(0.76)& \textbf{0.55}(0.74)& 189.94(3.5E124) & -\\
         & $200 \times 200$ & 0.79(0.69)& 2.00(0.77)& \textbf{0.54}(0.73)& 242.44(4.1E122) & -\\
         \hline
         \multirow{4}{*}{L2}& $20 \times 20$ & 2.95(0.82)& 4.92(0.63)& \textbf{0.90}(0.90)& 52.77(2.4E34) & 1.51(1.08)\\
         & $50 \times 50$ & 2.52(0.74)& 5.22(0.60)& \textbf{0.69}(0.82)& 103.02(4.3E109) & -\\
         & $100 \times 100$ & 2.29(0.69)& 4.89(0.57)& \textbf{0.61}(0.77)& 137.25(2.8E119) & -\\
         & $200 \times 200$ & 2.29(0.66)& 4.91(0.55)& \textbf{0.58}(0.75)& 176.24(1.2E124) & -\\
         \hline
    \end{tabular}
\end{table*}

Table \ref{table1} presents the average (across 100 simulated datasets) mean squared prediction errors (MSPE), along with corresponding average standard errors, obtained from fitting the models S1-S5 to data in test sets that varies according to censoring levels and grid sizes. Notably, under low levels of censoring (L1), the proposed model in scenario S3 yields comparable performance to situations where censored observations are excluded from analysis. However, ignoring spatial locations with censored observations entirely leads to unreliable estimates, particularly for the covariance parameters. Conversely, in instances of high data censoring (L2), the final model along with the imputation of censored observations (S3) outperforms all other models. It is noteworthy that the `CensSpatial' method also demonstrates relatively favorable performance for a grid size of $20 \times 20$, when the data-generating model is Mat\'ern GP; however, its computational inefficiency and inadequate scaling impeded our ability to apply the method to the higher-dimensional simulated datasets. In fact, \cite{ordonez2018geostatistical} showcased the efficacy of the `CensSpatial' algorithm through simulations involving only 50 and 200 spatial locations, clearly indicating its inadequacy regarding scalability.

\begin{table*}[h]
    \centering
    \caption{Median computation time (in minutes) corresponding to model fitting under the five different settings (S1-S5) to simulated test data from a Mat\'ern GP that varies with censoring levels L1 (low-censoring; 15\%) and L2 (high-censoring; 45\%), and grid sizes. The values in parenthesis represent the median absolute deviation for the corresponding computing times. Since model S5 (`CensSpatial') was infeasible for larger grids, table entries appear as `-'.}
    \label{table2}
    \begin{tabular}{ccccccc}
    \hline
         Censoring& Grid & \multicolumn{4}{c}{Computation Time (in minutes)}\\
         \cline{3-7}
         Level& size& S1 & S2 & S3 & S4 & S5\\ 
         \hline
         \multirow{4}{*}{L1}& $20 \times 20$ & 27.91(0.79)& 28.64(0.46)& 29.94(0.67)& 0.97(0.30) & 42.70(2.23) \\
         & $50 \times 50$ & 22.32(0.30)& 22.61(0.48)& 22.39(0.51)& 7.41(0.90) & -\\
         & $100 \times 100$ & 31.22(1.13)& 31.58(1.18)& 31.47(0.80)& 31.75(2.22) & -\\
         & $200 \times 200$ & 25.40(0.33)& 25.46(0.34)& 25.98(0.31)& 129.11(8.60) & -\\
         \hline
         \multirow{4}{*}{L2}& $20 \times 20$ & 26.47(0.50)& 28.82(0.48)& 29.68(0.92)& 3.99(0.26) & 80.40(3.53)\\
         & $50 \times 50$ & 21.67(0.45)& 22.70(0.56)& 22.78(0.22)& 25.82(0.92) & -\\
         & $100 \times 100$ & 29.28(1.05)& 30.18(1.41)& 31.06(1.14)& 104.14(2.69) & -\\
         & $200 \times 200$ & 24.74(0.51)& 25.47(0.27)& 26.56(0.36)& 423.26(8.90) & -\\
         \hline
    \end{tabular}
\end{table*}

Table \ref{table2} presents the median computation time corresponding to fitting the models S1-S5 to data in test sets that varies according to censoring levels and grid size. The computation times for S1, S2, and S3 are comparable, as they all employ the SPDE-approximated Mat\'ern GP, with runtime approximately proportional to the size of the INLA mesh used for process approximation (between 557 and 673 nodes for different datasets of different sizes). As anticipated, the runtime for the local likelihood approach utilizing Vecchia's approximation increases with larger grid sizes. As discussed earlier, the `CensSpatial' algorithm was infeasible for larger grids. The initial four methods, S1 - S4, were executed on SLURM clusters with one core per job and 8 GB RAM allocation. However, due to the current version of `CensSpatial' on CRAN being incompatible with the UNIX system, the algorithm was implemented on a personal Dell 7210 computer featuring 16 GB RAM, an Intel Core i5 dual-core processor, and a Windows 11 Enterprise 64-bit operating system.

\section{Application: California PFAS Data} 
\label{application}

\subsection{Analysis Plan and Hyperparameters}
We use the iterated log-transformed data (as described in Section \ref{data}) as our input to the proposed method. Since we have no covariates in this dataset, we use the coordinates of the locations (longitude, latitude) as covariates.  The hyperparameters for the priors are the same as mentioned in Section \ref{computation}. We fit a variogram model on the non-censored observations to obtain an initial set of parameter estimates for $\bm{\beta}, \tau$, $\phi$, and $r$. We run three chains with different starting values that are all close to the initial parameter estimates obtained by variogram fitting to allow for checking convergence and increasing the reliability of the model output. Each of the chains was run for $25,000$ iterations with the first $15,000$ samples discarded as burn-in. We thinned the post-burn-in samples by 5 to obtain $2,000$ samples from the posterior distribution of the parameters.

\subsection{Results}

\begin{figure*}
    \centering
    \includegraphics[height=0.48\linewidth]{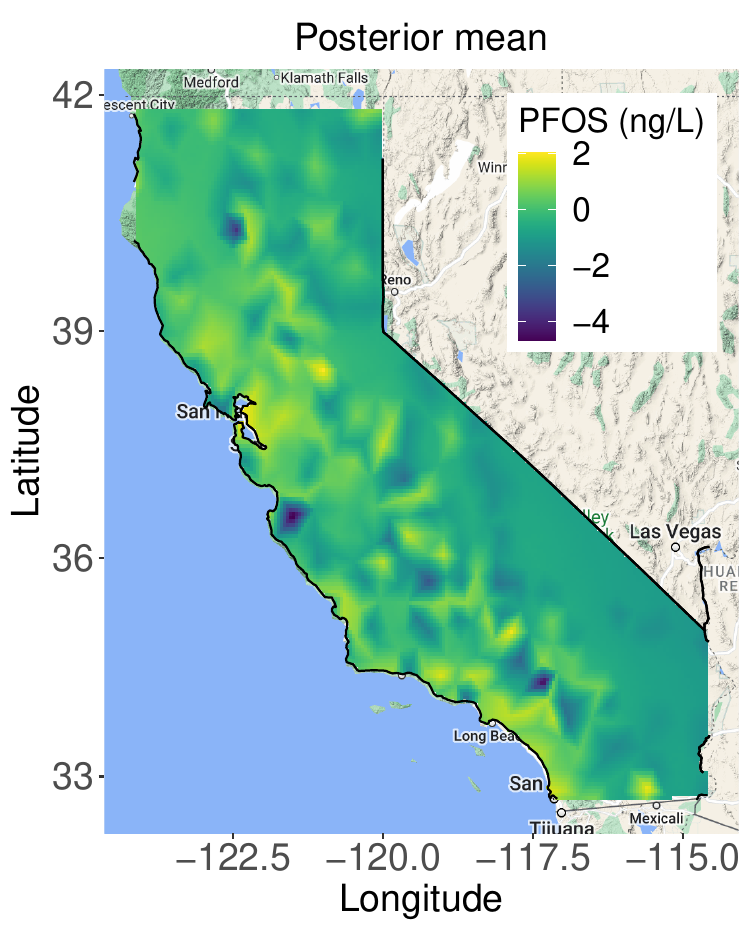}
    \includegraphics[height=0.48\linewidth]{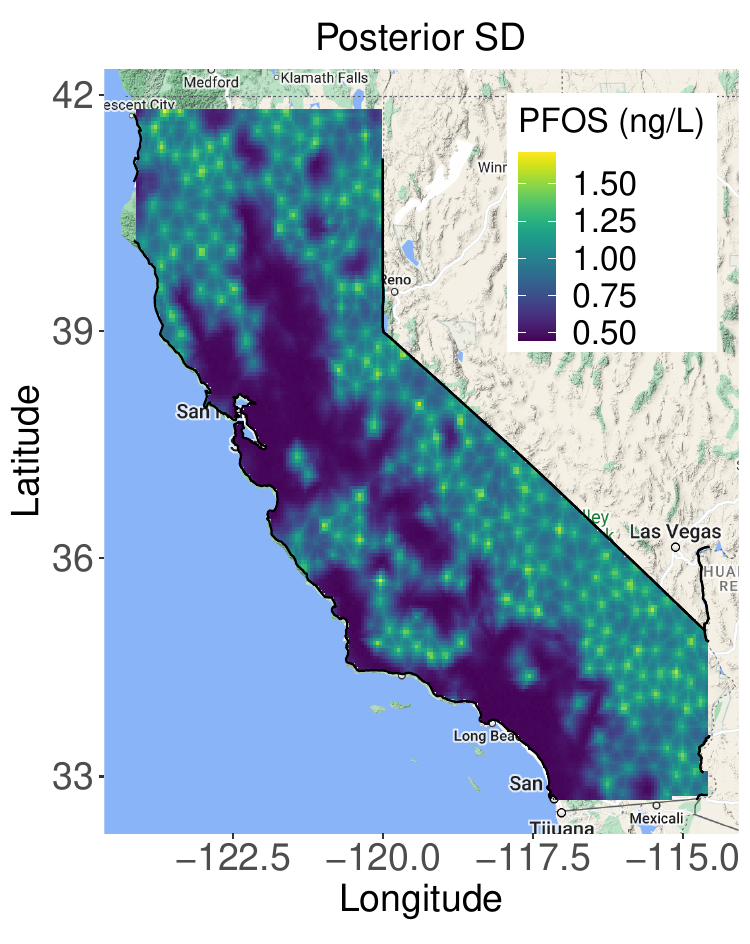}
    \caption{\emph{Left}: The predicted surface map for $g(\textrm{PFOS}) = \log(1 + \log(1 + \textrm{PFOS}))$ concentration in the state of California. \emph{Right}: The corresponding uncertainty estimates associated with the predictions for $g(\textrm{PFOS})$ across the pixels.}
    \label{pred_surf_sd}
\end{figure*}

For the observed data comprising $24,959$ locations and the prediction grid set at $0.1\degree \times 0.1\degree$, resulting in $405,893$ prediction locations across California, each of the three chains completed within approximately 62 minutes, with computations conducted on a SLURM cluster with an 8GB RAM allocation. We observe a reasonable well-mixing of the three chains. Each of the estimated covariate effects turned out to be significant (see Table \ref{data_res_tab}). The estimated spatial range is low ($\sim$7km). The prediction surface is smooth at places (left panel of \ref{pred_surf_sd}) with higher detailing around the regions with observed data. The prediction standard deviation is low towards the western parts where we have more observed data, but has an intriguing pattern on the east-southeastern parts (right panel of Figure \ref{pred_surf_sd}). We predict high values of PFOS concentration along the western part of the state, surpassing the EPA safety limit of approximately 0.45 in the transformed scale. Particularly notable are the elevated PFOS levels observed in and around Sonoma, Napa, Solano, Contra Costa, Alameda, San Francisco, and Santa Clara counties in central-west California, as well as in Los Angeles, Orange, and San Diego counties in southwest California. 

\begin{table}[ht]
\centering
\caption{Table of estimates and standard deviations corresponding to the parameters $\beta_0 - \beta_2$ denoting the intercept and the 2 covariates, respectively, $\phi$ (the spatial range), $\tau$ (the precision) and $r$ (ratio of partial sill to total variance).}
\label{data_res_tab}
\begin{tabular}{rrr}
  \hline
 & Estimates & Standard Deviation \\ 
  \hline
$\beta_0$ & $-22.04$ & 7.60 \\ 
  $\beta_1$ & $-0.23$ & 0.08 \\ 
  $\beta_2$ & $-0.14$ & 0.07 \\ 
  $\phi$ & 0.07 & 0.02 \\ 
  $\tau$ & 0.22 & 0.08 \\ 
  $r$ & 0.95 & 0.02 \\ 
   \hline
\end{tabular}
\end{table}

\section{Discussion}
\label{discussion}

We present a novel method to address the problem of modeling censored outcomes that are spatially correlated in big data settings. We observe that the proposed model scales nicely with an increased number of total observations and performs better than all other competing methods, even when nearly half of the observed data are censored. Despite being a fully Bayesian model, the runtime is moderate and better than the competing methods, highlighting its scalability, which combined with its demonstrated accuracy and precision makes this method an optimally efficient approximate method for modeling large spatial data in the presence of (left-) censoring. The real data analysis demonstrated this further as the model achieved satisfactory mixing of three chains for a large dataset in only an hour, producing sensible prediction surfaces and uncertainty quantification.

However, the data presents specific challenges during modeling, which may also be considered as limitations of the proposed method. The predicted surface in the left panel of Figure \ref{pred_surf_sd} is very smooth towards the east-southeast end of California. This is expected, as we have very few observations around that area to inform our spatial process. This, while being non-desirable, makes sense and is in line with what one would expect to happen for such a dataset. We can not hope to manufacture information in the absence of observations and we do not, reflecting the consistency of statistical principles being adhered to here in our analysis. The map of prediction uncertainty (right panel of Figure \ref{pred_surf_sd}) also reflects this. We have nearly zero uncertainty for the bulk of the region, where we observe numerous instances and have higher uncertainty whenever we move far from observations. Interestingly, we also notice a quilting pattern in the right panel of Figure \ref{pred_surf_sd}. This is a byproduct of the computation mesh considered here along with the lack of observations in the east-southeast region.

Further consideration are therefore needed for choosing the mesh and smoothness parameters for fitting the model. We explain our choice of mesh in Section \ref{data}. But this process is ad-hoc in nature, and a concrete workflow for selecting a mesh would be greatly beneficial for users. We consider this as a plausible future research direction. Another development on both the software and methodological fronts would be to include fractional smoothness parameters in the model, which is currently restricted to integer smoothness (we use $\nu = 1$ for all our analyses). One possible approach would be to use the fractional rational approximations to the SPDE model \citep{bolin2024covariance,bolin2020rational}. Combining the theory for fractional approximations with the software should render additional model flexibility, and more well-suited to real data applications. Further developments for multivariate extensions of the model that can model multiple spatial processes, both having a mix of censored and uncensored observations, simultaneously, are underway. 

Other than proposing a novel, scalable spatial model in its own right, the implications of this study extend beyond academic interest. By elucidating the spatial distribution of PFAS/PFOS contamination and its associated factors, we can inform targeted interventions, policy recommendations, and resource allocation to mitigate the impact of PFAS exposure on public health. Additionally, our research provides a framework, which can be adapted to analyze censored data in other environmental contexts, fostering a deeper understanding of complex contamination scenarios and enabling evidence-based decision-making.

\section*{Acknowledgement}
Arnab Hazra is partially supported by the Indian Institute of Technology Kanpur and Rice University collaborative research grant under Award No. DOIR/2023246. Dipankar Bandyopadhyay acknowledges funding support from grant R21DE031879 from the US National Institutes of Health.

\bibliographystyle{apalike}

\bibliography{biblio}
\end{document}